\catcode`\@=11                                   



\font\fiverm=cmr5                         
\font\fivemi=cmmi5                        
\font\fivesy=cmsy5                        
\font\fivebf=cmbx5                        

\skewchar\fivemi='177
\skewchar\fivesy='60


\font\sixrm=cmr6                          
\font\sixi=cmmi6                          
\font\sixsy=cmsy6                         
\font\sixbf=cmbx6                         

\skewchar\sixi='177
\skewchar\sixsy='60


\font\sevenrm=cmr7                        
\font\seveni=cmmi7                        
\font\sevensy=cmsy7                       
\font\sevenit=cmti7                       
\font\sevenbf=cmbx7                       

\skewchar\seveni='177
\skewchar\sevensy='60


\font\eightrm=cmr8                        
\font\eighti=cmmi8                        
\font\eightsy=cmsy8                       
\font\eightit=cmti8                       
\font\eightbf=cmbx8                       

\skewchar\eighti='177
\skewchar\eightsy='60


\font\ninei=cmmi9
\font\ninesy=cmsy9

\skewchar\ninei='177
\skewchar\ninesy='60


\font\tenrm=cmr10                         
\font\teni=cmmi10                         
\font\tensy=cmsy10                        
\font\tenex=cmex10                        
\font\tenit=cmti10                        
\font\tensl=cmsl10                        
\font\tenbf=cmbx10                        
\font\tentt=cmtt10                        
\font\tenss=cmss10                        
\font\tensc=cmcsc10                       
\font\tenbi=cmmib10                       

\skewchar\teni='177
\skewchar\tenbi='177
\skewchar\tensy='60

\def\tenpoint{\ifmmode\err@badsizechange\else
	\textfont0=\tenrm \scriptfont0=\sevenrm \scriptscriptfont0=\fiverm
	\textfont1=\teni  \scriptfont1=\seveni  \scriptscriptfont1=\fivemi
	\textfont2=\tensy \scriptfont2=\sevensy \scriptscriptfont2=\fivesy
	\textfont3=\tenex \scriptfont3=\tenex   \scriptscriptfont3=\tenex
	\textfont4=\tenit \scriptfont4=\sevenit \scriptscriptfont4=\sevenit
	\textfont5=\tensl
	\textfont6=\tenbf \scriptfont6=\sevenbf \scriptscriptfont6=\fivebf
	\textfont7=\tentt
	\textfont8=\tenbi \scriptfont8=\seveni  \scriptscriptfont8=\fivemi
	\def\rm{\tenrm\fam=0 }%
	\def\it{\tenit\fam=4 }%
	\def\sl{\tensl\fam=5 }%
	\def\bf{\tenbf\fam=6 }%
	\def\tt{\tentt\fam=7 }%
	\def\ss{\tenss}%
	\def\sc{\tensc}%
	\def\bmit{\fam=8 }%
	\rm\setparameters\setbaselines\fi}


\font\twelverm=cmr12                      
\font\twelvei=cmmi12                      
\font\twelvesy=cmsy10       scaled\magstep1             
\font\twelveex=cmex10       scaled\magstep1             
\font\twelveit=cmti12                            
\font\twelvesl=cmsl12                            
\font\twelvebf=cmbx12                            
\font\twelvett=cmtt12                            
\font\twelvess=cmss12                            
\font\twelvesc=cmcsc10      scaled\magstep1             
\font\twelvebi=cmmib10      scaled\magstep1             

\skewchar\twelvei='177
\skewchar\twelvebi='177
\skewchar\twelvesy='60

\def\twelvepoint{\ifmmode\err@badsizechange\else
	\textfont0=\twelverm \scriptfont0=\eightrm \scriptscriptfont0=\sixrm
	\textfont1=\twelvei  \scriptfont1=\eighti  \scriptscriptfont1=\sixi
	\textfont2=\twelvesy \scriptfont2=\eightsy \scriptscriptfont2=\sixsy
	\textfont3=\twelveex \scriptfont3=\tenex   \scriptscriptfont3=\tenex
	\textfont4=\twelveit \scriptfont4=\eightit \scriptscriptfont4=\sevenit
	\textfont5=\twelvesl
	\textfont6=\twelvebf \scriptfont6=\eightbf \scriptscriptfont6=\sixbf
	\textfont7=\twelvett
	\textfont8=\twelvebi \scriptfont8=\eighti  \scriptscriptfont8=\sixi
	\def\rm{\twelverm\fam=0 }%
	\def\it{\twelveit\fam=4 }%
	\def\sl{\twelvesl\fam=5 }%
	\def\bf{\twelvebf\fam=6 }%
	\def\tt{\twelvett\fam=7 }%
	\def\ss{\twelvess}%
	\def\sc{\twelvesc}%
	\def\bmit{\fam=8 }%
	\rm\setparameters\setbaselines\fi}


\font\fourteenrm=cmr12      scaled\magstep1             
\font\fourteeni=cmmi12      scaled\magstep1             
\font\fourteensy=cmsy10     scaled\magstep2             
\font\fourteenex=cmex10     scaled\magstep2             
\font\fourteenit=cmti12     scaled\magstep1             
\font\fourteensl=cmsl12     scaled\magstep1             
\font\fourteenbf=cmbx12     scaled\magstep1             
\font\fourteentt=cmtt12     scaled\magstep1             
\font\fourteenss=cmss12     scaled\magstep1             
\font\fourteensc=cmcsc10 scaled\magstep2  
\font\fourteenbi=cmmib10 scaled\magstep2  

\skewchar\fourteeni='177
\skewchar\fourteenbi='177
\skewchar\fourteensy='60

\def\fourteenpoint{\ifmmode\err@badsizechange\else
	\textfont0=\fourteenrm \scriptfont0=\tenrm \scriptscriptfont0=\sevenrm
	\textfont1=\fourteeni  \scriptfont1=\teni  \scriptscriptfont1=\seveni
	\textfont2=\fourteensy \scriptfont2=\tensy \scriptscriptfont2=\sevensy
	\textfont3=\fourteenex \scriptfont3=\tenex \scriptscriptfont3=\tenex
	\textfont4=\fourteenit \scriptfont4=\tenit \scriptscriptfont4=\sevenit
	\textfont5=\fourteensl
	\textfont6=\fourteenbf \scriptfont6=\tenbf \scriptscriptfont6=\sevenbf
	\textfont7=\fourteentt
	\textfont8=\fourteenbi \scriptfont8=\tenbi \scriptscriptfont8=\seveni
	\def\rm{\fourteenrm\fam=0 }%
	\def\it{\fourteenit\fam=4 }%
	\def\sl{\fourteensl\fam=5 }%
	\def\bf{\fourteenbf\fam=6 }%
	\def\tt{\fourteentt\fam=7}%
	\def\ss{\fourteenss}%
	\def\sc{\fourteensc}%
	\def\bmit{\fam=8 }%
	\rm\setparameters\setbaselines\fi}


\font\seventeenrm=cmr10 scaled\magstep3          


\newdimen\rp@
\newcount\@basestretchnum
\newskip\@baseskip
\newskip\headskip
\newskip\footskip


\def\setparameters{\rp@=.1em
	\headskip=24\rp@
	\footskip=\headskip
	\delimitershortfall=5\rp@
	\nulldelimiterspace=1.2\rp@
	\scriptspace=0.5\rp@
	\abovedisplayskip=10\rp@ plus3\rp@ minus5\rp@
	\belowdisplayskip=10\rp@ plus3\rp@ minus5\rp@
	\abovedisplayshortskip=5\rp@ plus2\rp@ minus4\rp@
	\belowdisplayshortskip=10\rp@ plus3\rp@ minus5\rp@
	\normallineskip=\rp@
	\lineskip=\normallineskip
	\normallineskiplimit=0pt
	\lineskiplimit=\normallineskiplimit
	\jot=3\rp@
	\setbox0=\hbox{\the\textfont3 B}\p@renwd=\wd0
	\skip\footins=12\rp@ plus3\rp@ minus3\rp@
	\skip\topins=0pt plus0pt minus0pt}


\def\setbaselines{\maxdepth=4\rp@\baselinestretch=\@basestretchnum}


\def\baselinestretch{\afterassignment\@basestretch\@basestretchnum}
\def\@basestretch{%
	\@baseskip=12\rp@ \divide\@baseskip by1000
	\normalbaselineskip=\@basestretchnum\@baseskip
	\baselineskip=\normalbaselineskip
	\bigskipamount=\the\baselineskip
		plus.25\baselineskip minus.25\baselineskip
	\medskipamount=.5\baselineskip
		plus.125\baselineskip minus.125\baselineskip
	\smallskipamount=.25\baselineskip
		plus.0625\baselineskip minus.0625\baselineskip
	\setbox\strutbox=\hbox{\vrule height.708\baselineskip
		depth.292\baselineskip width0pt }}



\def\makeheadline{\vbox to0pt{\baselinestretch=1000
	\vskip-\headskip \vskip1.5pt
	\line{\vbox to\ht\strutbox{}\the\headline}\vss}\nointerlineskip}

\def\makefootline{\baselineskip=\footskip\line{\the\footline}}

\def\big#1{{\hbox{$\left#1\vbox to8.5\rp@ {}\right.\n@space$}}}
\def\Big#1{{\hbox{$\left#1\vbox to11.5\rp@ {}\right.\n@space$}}}
\def\bigg#1{{\hbox{$\left#1\vbox to14.5\rp@ {}\right.\n@space$}}}
\def\Bigg#1{{\hbox{$\left#1\vbox to17.5\rp@ {}\right.\n@space$}}}


\mathchardef\alpha="710B
\mathchardef\beta="710C
\mathchardef\gamma="710D
\mathchardef\delta="710E
\mathchardef\epsilon="710F
\mathchardef\zeta="7110
\mathchardef\eta="7111
\mathchardef\theta="7112
\mathchardef\iota="7113
\mathchardef\kappa="7114
\mathchardef\lambda="7115
\mathchardef\mu="7116
\mathchardef\nu="7117
\mathchardef\xi="7118
\mathchardef\pi="7119
\mathchardef\rho="711A
\mathchardef\sigma="711B
\mathchardef\tau="711C
\mathchardef\upsilon="711D
\mathchardef\phi="711E
\mathchardef\chi="711F
\mathchardef\psi="7120
\mathchardef\omega="7121
\mathchardef\varepsilon="7122
\mathchardef\vartheta="7123
\mathchardef\varpi="7124
\mathchardef\varrho="7125
\mathchardef\varsigma="7126
\mathchardef\varphi="7127
\mathchardef\imath="717B
\mathchardef\jmath="717C
\mathchardef\ell="7160
\mathchardef\wp="717D
\mathchardef\partial="7140
\mathchardef\flat="715B
\mathchardef\natural="715C
\mathchardef\sharp="715D


\def\err@badsizechange{%
	\immediate\write16{--> Size change not allowed in math mode, ignored}}

\baselinestretch=1000
\tenpoint

\catcode`\@=12                                   
\catcode`\@=11
\expandafter\ifx\csname @iasmacros\endcsname\relax
	\global\let\@iasmacros=\par
\else  \immediate\write16{}
	\immediate\write16{Warning:}
	\immediate\write16{You have tried to input iasmacros more than once.}
	\immediate\write16{}
	\endinput
\fi
\catcode`\@=12


\def\rmb{\seventeenrm}

\def\singlespace{\baselineskip=\normalbaselineskip}
\def\halfspace{\baselineskip=1.5\normalbaselineskip}
\def\doublespace{\baselineskip=2\normalbaselineskip}


\def\AB{\bigskip\parindent=40pt
	 \centerline{\bf ABSTRACT}\medskip\halfspace\narrower}
\def\AE{\bigskip\nonarrower\doublespace}
\def\nonarrower{\advance\leftskip by-\parindent
	\advance\rightskip by-\parindent}


\def\boxit#1{\vbox{\hrule\hbox{\vrule\kern3pt
	\vbox{\kern3pt#1\kern3pt}\kern3pt\vrule}\hrule}}

\def\hence{\leavevmode\hbox{\bf .\raise5.5pt\hbox{.}.} }

\def\dalemb#1#2{{\vbox{\hrule height.#2pt
	\hbox{\vrule width.#2pt height#1pt \kern#1pt \vrule width.#2pt}
	\hrule height.#2pt}}}
\def\gtorder{\mathrel{\raise.3ex\hbox{$>$}\mkern-14mu
	      \lower0.6ex\hbox{$\sim$}}}
\def\ltorder{\mathrel{\raise.3ex\hbox{$<$}\mkern-14mu
	      \lower0.6ex\hbox{$\sim$}}}

\newdimen\fullhsize
\newbox\leftcolumn
\def\twoup{\hoffset=-.5in \voffset=-.25in
  \hsize=4.75in \fullhsize=10in \vsize=6.9in
  \def\fullline{\hbox to\fullhsize}
  \let\lr=L
  \output={\if L\lr
	 \global\setbox\leftcolumn=\columnbox\global\let\lr=R \advancepageno
      \else \doubleformat \global\let\lr=L\fi
    \ifnum\outputpenalty>-20000 \else\dosupereject\fi}
  \def\doubleformat{\shipout\vbox{
    \fullline{\box\leftcolumn\hfil\columnbox}\advancepageno}}
  \def\columnbox{\leftline{\vbox{\makeheadline\pagebody\makefootline}}}
  \tolerance=1000 }

\twelvepoint
\doublespace
{\nopagenumbers{

\rightline{~~~November, 2006}
\bigskip\bigskip
\centerline{\rmb Vacuum Birefringence in a}
\centerline{\rmb Rotating Magnetic Field}
\medskip
\centerline{\it Stephen L. Adler}
\centerline{\bf Institute for Advanced Study}
\centerline{\bf Princeton, NJ 08540}
\medskip
\bigskip\bigskip
\leftline{\it Send correspondence to:}
\medskip
{\singlespace\leftline{Stephen L. Adler}
\leftline{Institute for Advanced Study}
\leftline{Einstein Drive, Princeton, NJ 08540}
\leftline{Phone 609-734-8051; FAX 609-924-8399; 
email adler@ias.edu}}
\bigskip\bigskip
}}
\vfill\eject
\pageno=2
\AB
We calculate the vacuum polarization-induced 
ellipticity acquired by a linearly polarized laser beam  of angular 
frequency $\bar \omega$ 
on traversing a region containing a transverse 
magnetic field rotating with a small 
angular velocity $\Omega$ around the beam axis.   
The transmitted beam contains the fundamental 
frequency $\bar \omega$ and weak sidebands of frequency 
$\bar \omega\pm 2\Omega$, 
but no other sidebands.  To first order in small quantities, the 
ellipticity acquired by the  
transmitted beam is independent of $\Omega$, and is 
the same as would be calculated in the  
approximation of regarding the magnetic field as fixed at its instantaneous 
angular orientation,  using the standard vacuum birefringence formulas for 
a static magnetic field.   Also to first order, there is no rotation of 
the polarization plane of the transmitted beam.  Analogous statements   
hold when the magnetic field strength is slowly varying in time.    
\AE
\bigskip\bigskip
\vfill\eject
\pageno=3
\centerline{{\bf 1.~~Introduction}}
\bigskip
There has recently been renewed interest in trying to measure the 
birefringence of the vacuum in a strong magnetic field predicted by 
the Heisenberg--Euler effective Lagrangian [1,2,3,4] for 
quantum electrodynamics (QED).  
Such a measurement was a primary motivation for the PVLAS experiment [5], 
which however has reported a signal about $10^4$ times larger than 
expected from QED.  In a recent article [6], a claim was made that 
just the kinematics of rotation of the magnetic field used in the PVLAS 
experiment, through the generation of multiple sidebands to the transmitted 
laser beam, could explain the PVLAS result.  However, examination 
of ref [6] and a related paper [7] suggests several problems [8].  
First, the 
authors have overestimated the strength of the vacuum birefringence 
coefficient by a factor of over 1000.   Second, since on  bases rotating 
with the magnetic field, 
the vacuum as modified by the magnetic field still acts  
on an incident electromagnetic wave as a time-independent 
linear medium, the presence 
of an infinite sequence of sidebands (as would be obtained from a nonlinear 
medium) is not expected; the only sidebands should be those associated    
with transformations between fixed laboratory bases and the rotating bases. 

Nonetheless, the authors of ref [6] have called 
 attention to an interesting, and as it turns 
out quite nontrivial, problem in wave propagation.  This paper is devoted 
to formulating and solving this problem to leading order, using standard 
scattering theory and perturbation theory methods.   
Our conclusion is that to leading order 
in small quantities, the method used by the PVLAS group to calculate the 
birefringence-induced ellipticity yields the correct answer, which is  
independent of the rotation angular velocity  $\Omega$. 

To set the stage, we start from the lowest order formulas describing 
electromagnetic wave propagation in an external magnetic field, as 
deduced from the Heisenberg--Euler effective Lagrangian.  With repeated 
indices summed, the $\vec d$ and $\vec h$ fields are obtained from the 
$\vec e$ and $\vec b$ fields by 
$$\eqalign{
d_s=&\epsilon_{st}e_t  ~~~,\cr
h_s=&\mu^{-1}_{st}b_t~~~.\cr
}\eqno(1a)$$
The polarization tensors $\epsilon_{st}$ and $\mu^{-1}_{st}$ arising 
from the fourth order box diagram are given by 
$$\eqalign{
\epsilon_{st}=&\delta_{st}(1-2\xi)+7\xi \hat B_s(t) \hat B_t(t)~~~,\cr
\mu^{-1}_{st}=&\delta_{st}(1-2\xi)-4\xi \hat B_s(t) \hat B_t(t)~~~,\cr
}\eqno(1b)$$
with $\hat B_s(t)$ a unit vector along the magnetic field, which we take 
to have fixed magnitude $B$.   
With $B$ measured in unrationalized units, and setting  $\hbar=c=1$,  
the parameter  $\xi$ governing the strength  
of the QED vacuum polarization effect is given by  
$$\xi={\alpha^2 B^2 \over 45 \pi m^4} ~~~,\eqno(1c)$$
where  $\alpha\simeq 1/137.04$ is the fine structure constant and $m$ is the 
electron mass.  Finally, in terms of $\vec d, \vec h, \vec e$, 
and $\vec b$ the  
Maxwell equations take the form 
$$\eqalign{
\vec \nabla \cdot \vec d=&\vec \nabla \cdot \vec b =0~~~,\cr
\vec \nabla \times \vec e =& -{\partial \vec b \over \partial t}~,~~~ 
\vec \nabla \times \vec h = {\partial \vec d \over \partial t}~~~.\cr
}\eqno(1d)$$
Equations (1a-d) are the fundamental physical equations governing the 
wave propagation problem under study.   When the magnetic field orientation 
$\hat B(t)$ is time-independent, analysis of this problem shows that 
the vacuum becomes birefringent.  Specializing to the case of a wave-vector 
$\vec k$ perpendicular to $\hat B$, the polarization state with $\vec e$ 
perpendicular to $\hat B$ propagates with index of refraction 
$k/\omega=1+2\xi$, while the polarization state with $\vec e$ parallel 
to $\hat B$ propagates with index of refraction $k/\omega=1+{7\over 2}\xi$ 
(with $k=|\vec k|$). 
We will denote the difference between the larger and smaller refractive 
indices by $\Delta n={3\over 2} \xi$.  

To finish specifying the problem, we shall assume that the linearly 
polarized laser beam 
is propagating upwards along the $z$ axis, with field strengths (expressed  
 in terms of $(x,y,z)$ components) given by 
$$\eqalign{
\vec e=&\vec d = e^{i \bar \omega(z-t)}(\cos \theta,\sin \theta, 0)~~~,\cr 
\vec b=&\vec h = e^{i \bar \omega(z-t)}(-\sin \theta,\cos \theta, 0)~~~.\cr
}\eqno(2a)$$
We assume that the the uniform magnetic field is confined to the 
region $0\leq z \leq L$, and is oriented perpendicular to the $z$ axis, 
and rotates uniformly with angular velocity $\Omega$, according to 
$$\hat B(t)=(\cos \Omega t, \sin \Omega t,0) ~~~,\eqno(2b)$$ 
from which we learn that (with the prime denoting time differentiation) 
$$ \hat B^{\prime}(t)= \Omega (-\sin \Omega t, \cos \Omega t,0)  
=\Omega \hat z \times \hat B~~.\eqno(2c)$$   
The problem is then to 
calculate the upward moving transmitted wave emerging at $z=L$.

Before proceeding with details, we outline our basic strategy.  The first 
observation is that since 
Eq.~(2c) relates the time derivative of $\hat B$ 
to $\hat z \times \hat B$ with a time-independent coefficient, the 
wave propagation problem in the region $0\leq z \leq L$ will involve 
differential equations with time-independent coefficients when referred 
to the orthonormal rotating bases $\hat B$ and $\hat z \times \hat B$.  
These differential 
equations can then be solved by a standard traveling wave Ansatz; this 
calculation of the propagation modes in the rotating 
magnetic field ``medium'' is 
carried out in Sec. 2.  One then has to transform the incident wave to 
the rotating bases, and do a matching of incident, transmitted, and 
reflected waves at the medium boundaries $z=0$ and $z=L$.  This 
calculation is 
carried out in Sec. 3.  The calculation is greatly facilitated by working 
to leading order in the small quantities $\xi$,  $\Omega/\bar \omega$, 
and $\Omega L$, that is, we regard both $\xi$ and $\Omega$ as small.   
Since each reflected wave is reduced in strength by $O(\xi)$, this 
allows us to neglect multiple reflections, such as a wave reflected back 
from $z=L$, and then reflected forward from $z=0$, and finally  emerging 
at $z=L$, which makes a contribution of order $\xi^2$ to the transmitted   
wave.  When multiple reflections are neglected, one can solve the wave 
propagation problem (to leading order) by considering independent 
matching problems at $z=0$ and $z=L$.  Moreover, 
it then suffices to compute the phases of the various waves to first order 
in small quantities, but the corresponding transmitted wave amplitudes 
only to zeroth order. The reason this approximation is adequate 
is that the transmission coefficients at the two boundaries are independent 
of $L$  when multiple reflections are neglected, and so first order terms 
in the transmission coefficients can contribute only an $L$-independent 
term to the polarization parameters of the emerging wave at $z=L$.  
However, when 
$L$=0  no birefringent medium is traversed, and so the ellipticity 
of the emerging wave is zero; hence the first order corrections to 
the wave coefficients must make a vanishing contribution to the ellipticity,     
and similarly, to the rotation of the polarization axis.  
The calculation of the ellipticity and polarization axis, from our 
formulas for the emerging 
waves, is given in Sec. 4.  An alternative calculation method, based on 
a perturbative expansion in $\xi$, is given in Sec. 5, which gives results 
identical to the wave-matching calculation of Sec. 3 for the rotating 
field case, and also applies 
to the case when the magnetic field strength is time dependent. 
A brief discussion of our results and their experimental implications 
is given in Sec. 6.  
\bigskip
\centerline{{\bf 2.~~Electromagnetic wave eigenmodes in a rotating 
magnetic field}}
\bigskip
We begin by calculating the electromagnetic wave eigenmodes in the region 
containing the rotating magnetic field.  We write the $\vec e$ and 
$\vec b$ fields in terms 
of components along the rotating bases $\hat B$ and $\hat z \times \hat B$,
$$\eqalign{
\vec e=&[E_1(t) \hat B + E_2(t) \hat z \times \hat B] e^{ikz}~~~,\cr   
\vec b=&[B_1(t) \hat B + B_2(t) \hat z \times \hat B] e^{ikz}~~~.\cr   
}\eqno(3a) $$
Substituting these into Eq.~(1a), and using Eq.~(1b), we get the 
corresponding expressions for the $\vec d$ and $\vec h$ fields, 
$$\eqalign{
\vec d=&[(1+5\xi)E_1(t) \hat B +(1-2\xi) E_2(t) 
\hat z \times \hat B] e^{ikz}~~~,\cr   
\vec h=&[(1-6\xi)B_1(t) \hat B + (1-2\xi)B_2(t) 
\hat z \times \hat B] e^{ikz}~~~.\cr   
}\eqno(3b) $$
Substituting Eqs.~(3a,b) into the Maxwell equations of Eq.~(1d), and using 
Eq.~(2c), we get a set of coupled equations for the coefficient functions 
$E_{1,2}(t)$ and $B_{1,2}(t)$, 
$$\eqalign{
ikE_1(t)=&-\Omega B_1(t) -B_2^{\prime}(t)~~~,\cr
-ikE_2(t)=&-B_1^{\prime}(t)+\Omega B_2(t)~~~.\cr
ik(1-6\xi)B_1(t)=&(1+5\xi)\Omega E_1(t)+(1-2\xi)E_2^{\prime}(t)~~~,\cr 
-ik(1-2\xi)B_2(t)=&(1+5\xi)E_1^{\prime}(t)-(1-2\xi)\Omega E_2(t)~~~.\cr
}\eqno(4a)$$
Since this set of equations has time independent coefficients, it can be 
solved by making an exponential Ansatz, 
$$E_{1,2}(t)=E_{1,2}e^{-i\omega t}~,~~~B_{1,2}(t)=B_{1,2}e^{-i\omega t}
~~~.\eqno(4b)$$
Substituting Eq.~(4b) into Eq.~(4a), and simplifying by neglecting 
second order small terms proportional to $\xi^2$ and $\xi \Omega$, 
we get the coupled equations for the constant amplitudes $E_{1,2}$ and 
$B_{1,2}$, 
$$\eqalign{
kE_1=&\omega B_2+\Omega\, iB_1~~~,\cr
k\, iE_2=&-\omega iB_1 -\Omega B_2~~~,\cr
k\,iB_1=&-\omega(1+4\xi)\, iE_2 +\Omega E_1~~~,\cr
kB_2=&\omega(1+7\xi) E_1-\Omega\, iE_2~~~.
}\eqno(4c)$$
We note at this point that we have ignored effects arising from  
the electric field induced by the rotating magnetic field; these are 
of order $\xi \Omega$ at least, and so are second order in the 
small parameters of the problem.  Similarly, other relativistic 
effects associated with electrodynamics in 
rotating frames, on which there is a 
substantial literature, are not relevant for our analysis.   

Since Eq.~(4c) is a homogeneous set of linear equations, a solution is 
possible only when the determinant vanishes, which yields a quartic 
equation giving the dispersion relation of the wave,  
$$[k^2-\omega^2(1+7\xi)][k^2-\omega^2(1+4\xi)]
-2\Omega^2(\omega^2+k^2) + \Omega^4=0~~~.\eqno(5a)$$
This can be solved to give two solutions $k_{\pm}$ which describe  upward 
propagating waves, and two solutions $-k_{\pm}$ which describe  downward 
propagating waves,  with $k_{\pm}$ given by 
$$\eqalign{
k_{\pm}=&\omega\pm\Omega +{1\over 2} \omega \sigma_{\pm}~~~,\cr
\sigma_{\pm}\equiv&\sigma_{\pm}(\omega)={11\over 2} 
\xi \pm\big[\big((2\Omega/\omega)^2+(3\xi/2)^2\big)^{1\over 2} 
-2\Omega/\omega\big]~~~,\cr
}\eqno(5b)$$
so that the corresponding refractive indices are given by 
$$k_{\pm}/\omega = n_{\pm}=1 \pm {\Omega \over \omega} + 
{1\over 2} \sigma_{\pm}~~~.\eqno(5c)$$
Note that $\sigma_{\pm}$ is first order in small quantities, and is uniformly 
of order $\xi$ or smaller irrespective of the relative sizes of 
$\xi$ and $\Omega/\omega$.  Also, we note that the terms of order 
$\xi \Omega$ that we have dropped make a contribution inside the square root 
of order $\xi (\Omega/\omega)^2$ (because the dispersion relation must 
be an even function of $\Omega$, as we have verified by 
explicit computation), and so is a higher order effect relative to the 
quadratic terms that we retained inside the square root.  

Solving for the ratios of the field coefficients to $E_1$, we get for the 
case of upward propagating waves with 
oscillatory behavior $e^{i(n_{\pm}z-t)\omega}$
$$\eqalign{
{B_2 \over E_1}=&1 +\beta_{\pm}~~~, \cr
{iE_2\over E_1}=&{\mp}1 + \alpha_{\pm}~~~,\cr
{iB_1\over E_1}=&\pm 1+\gamma_{\pm}~~~,\cr
}\eqno(6a)$$
with the  quantities $\alpha_{\pm},\,\beta_{\pm},\,\gamma_{\pm}$,  
which all vanish when $\xi$ vanishes, given by 
$$\eqalign{
\beta_{\pm}=&{7\over 2}\xi~~~,\cr
\alpha_{\pm}=&{\omega \over 2 \Omega} (1-2\xi)(7\xi-\sigma_{\pm}) \pm 2 \xi
~~~\cr
\simeq&{\omega \over 2 \Omega} (7\xi-\sigma_{\pm})~~~,\cr
\gamma_{\pm}=&-{\omega\over 2\Omega}(2+2\xi-n_{\pm})(7\xi-\sigma_{\pm})
\pm (2\xi-{1\over 2} \sigma_{\pm})~~~\cr
\simeq&-\alpha_{\pm}~~~,\cr
}\eqno(6b)$$
where the first lines of the expressions for $\alpha_{\pm}$ 
and $\gamma_{\pm}$ are accurate to first order in small quantities, and 
the second lines are accurate to leading zeroth order.  For the case of 
downward propagating waves with 
oscillatory behavior $e^{i(-n_{\pm}z-t)\omega}$, one simply reverses 
the sign of $B_2/E_1$ and $iB_1/E_1$, while keeping $iE_2/E_1$ the same 
as in Eqs.~(6a,b).   This completes the analysis of the electromagnetic 
wave eigenmodes in the region $0\leq z\leq L$ containing the rotating 
magnetic field $B$.  
\bigskip
\centerline{\bf 3.~~Wave matching to get the transmission coefficient}
\bigskip
Now that we know the form of propagating waves in the magnetic field 
region, we turn to the problem of determining the emerging waves at 
$z\geq L$ produced by the laser beam incident from below $z=0$.  As noted 
in Sec. 1, when doubly reflected waves are ignored, we can solve this 
problem by treating sequentially the wave matching at $z=0$, followed by 
the wave matching at $z=L$.  The first step is to rewrite the incident 
wave at $z=0$ on the rotating basis, giving 
$$\eqalign{
\vec e|_{z=0}=&e^{-i\bar \omega t}[\cos(\theta-\Omega t) \hat B 
+ \sin(\theta-\Omega t) \hat z\times \hat B]~~~,\cr
\vec b|_{z=0}=&e^{-i\bar \omega t}[-\sin(\theta-\Omega t) \hat B 
+ \cos(\theta-\Omega t) \hat z\times \hat B]~~~,\cr
}\eqno(7)$$
which shows that all the components of the incident wave on the 
rotating bases have the time dependence $e^{-i(\bar \omega +\Omega)t}$ or 
$e^{-i(\bar \omega -\Omega)t}$.  Thus to do the match, we need a reflected 
wave at $z\leq0$ with four coefficients (two for the reflected wave  
with frequency $\bar\omega+\Omega$, and two for the reflected wave with 
frequency $\bar\omega-\Omega$), and a transmitted wave with four coefficients  
(one each for the $\pm$ eigenmodes with frequency $\omega=\bar \omega 
+ \Omega$,  
and one each for the $\pm$ eigenmodes with frequency $\omega=\bar \omega -
\Omega$).  Continuity of the $\vec e$ and $\vec h$ fields across the junction 
at $z=0$ gives eight matching conditions (there are two field components 
parallel to the boundary for each of the two fields to be matched, and 
two frequency components for each).  Solving the wave matching equations 
shows, as expected, that the reflected wave amplitudes are of order $\xi$, 
so to get the transmitted wave amplitudes to zeroth order it suffices to 
ignore the reflected waves.  Similarly, at $z=L$ there are also eight 
unknown coefficients, four for the transmitted wave (two  for   
frequency $\bar \omega +\Omega$ and two for frequency  
$\bar \omega -\Omega$), and four for 
the waves of the two frequencies reflected back into the magnetic 
field region $z\leq L$.  Again there are eight matching conditions, and 
solving for the reflected waves shows that they are again of order $\xi$ 
in magnitude, and so again the leading order coefficients of the transmitted 
wave can be obtained by ignoring the reflected waves.  

Thus, the transmitted 
wave on the rotating basis, to leading order in the wave amplitudes, is 
simply the result of compounding transmission at $z=0$, followed by  
propagation 
from $z=0$ to $z=L$, and then followed by transmission at $z=L$.  
The final step 
is to convert back from the rotating bases to fixed laboratory bases.  
This yields three distinct emerging wave components, with 
frequencies $\bar \omega$, $\bar \omega+2\Omega$ and $\bar \omega-2\Omega$.   
We dispense with all of the intermediate algebra, and simply present the 
final result for the transmitted wave at $z=L$, which takes the 
form 
$$\eqalign{
\vec e|_{z=L}=&e^{i\bar \omega (L-t)}(X,Y,0)~~~,\cr
\vec b|_{z=L}=&e^{i\bar \omega (L-t)}(-Y,X,0)~~~,\cr
}\eqno(8a)$$
with the components $X,Y$ given in terms of quantities $A,B,C,D$ by 
$$ \eqalign{
X=&{1\over 2}(C+D+Ae^{2i\Omega t} +Be^{-2i\Omega t})~~~,\cr 
Y=&{i\over 2}(-C+D-Ae^{2i\Omega t} +Be^{-2i\Omega t})~~~.\cr 
}\eqno(8b)$$
Introducing the definitions 
$$\eqalign{
\phi_{\pm}\equiv&{1\over 2} \bar \omega \sigma_{\pm}(\bar \omega)~~~,\cr 
a_{\pm}\equiv&{\bar \omega \over 4 \Omega}[7\xi-\sigma_{\pm}(\bar \omega)]~~~,\cr
}\eqno(8c)$$
the auxiliary quantities $A,B,C,D$ are given by the following formulas, 
$$\eqalign{    
A=&(e^{i\phi_+L}-e^{i(-2\Omega +\phi_-)L})e^{-i\theta} 
{(1+a_-)a_+\over 1+a_--a_+}~~~,\cr 
B=&(e^{i(2\Omega+\phi_+)L}-e^{i\phi_-L}) e^{i\theta} 
{(1-a_+)a_- \over  1+a_--a_+}~~~,\cr  
C=&e^{i\theta +i(2\Omega+\phi_+)L} {a_+a_-\over 1+a_--a_+}
 +e^{i\theta+i\phi_-L} {(1-a_+)(1+a_-) \over  1+a_--a_+}~~~,\cr  
D=&e^{-i\theta +i(-2\Omega+\phi_-)L} {a_+a_-\over 1+a_--a_+}
 +e^{-i\theta+i\phi_+L} {(1-a_+)(1+a_-) \over  1+a_--a_+}~~~,\cr  
}\eqno(9a)$$
These formulas are accurate to first order in small quantities in the phases, 
and to zeroth order in small quantities in the real amplitudes multiplying 
the phases.  As we stressed in Sec.~1, first order corrections to 
the real amplitudes are independent  
of $L$, and so can make no contribution to physical attributes, such as 
the ellipticity, of the emerging wave.  
Expanding the exponentials through first order in   
small quantities, we get the following expressions for $A,B,C,D$ 
\big(with $\sigma_{\pm}\equiv\sigma_{\pm}(\bar \omega)$\big), 
$$\eqalign{
A=&i\bar \omega L\left[ {2\Omega\over\bar\omega}+{1\over 2} (\sigma_+  
-\sigma_-)\right] e^{-i\theta} 
{(1+a_-)a_+\over 1+a_--a_+}~~~,\cr 
B=&i\bar \omega L\left[ {2\Omega\over\bar\omega}+{1\over 2} (\sigma_+  
-\sigma_-)\right] e^{i\theta} 
{(1-a_+)a_- \over  1+a_--a_+}~~~,\cr  
C=&\left[1+i\bar\omega L 
\left({2\Omega\over \bar \omega}+{1\over 2} \sigma_+\right) 
 {a_+a_-\over 1+a_--a_+} 
 +i\bar \omega L {1\over 2} \sigma_-  {(1-a_+)(1+a_-) \over  1+a_--a_+} 
 \right] e^{i\theta}~~~,\cr
D=&\left[1+i\bar \omega L {1\over 2} \sigma_+  
{(1-a_+)(1+a_-) \over  1+a_--a_+} 
+i\bar \omega L \left(-{2\Omega\over \bar \omega} 
+ {1\over 2} \sigma_- \right)  {a_+a_-\over 1+a_--a_+} \right]e^{-i\theta}
~~~.\cr
}\eqno(9b)$$
These equations can be greatly simplified by using the following three 
algebraic identities obeyed by the quantities $a_{\pm}$, 
$$\eqalign{
a_+-a_-+2a_+a_-=&0~~~,\cr
{(1-a_+)(1+a_-)\over 1+a_--a_+}=&1-{a_+a_-\over 1+a_--a_+}~~~,\cr
1+a_--a_+=&1+{\bar \omega \over 4 \Omega}(\sigma_+-\sigma_-)~~~.\cr
}\eqno(10a)$$
We then get the remarkably simple formulas 
$$\eqalign{
X=&\cos \theta (1+ {11\over 4} i\bar \omega L \xi) 
+ {3\over 4}i \bar \omega L \xi \cos(2\Omega t-\theta)~~~,\cr
Y=&\sin \theta (1+ {11\over 4} i\bar \omega L \xi) 
+ {3\over 4}i \bar \omega L \xi \sin(2\Omega t-\theta)~~~,\cr
}\eqno(10b)$$
which together with Eq.~(8a) are our final result for 
the wave transmission problem.   They 
will be used in the next section to calculate the ellipticity and major 
axis orientation of the emerging wave.

The vacuum with magnetic field is a specific instance of a generic 
weakly birefringent medium which has orthogonal polarization eigenmodes 
with refractive indices $n_{\parallel}$,  $n_{\perp}$.  Hence, with the 
replacements ${7\over 2} \xi \to n_{\parallel}-1$, 
$2\xi \to n_{\perp}-1$, ${11\over 2} \xi \to n_{\parallel}+n_{\perp}-2$, 
and ${3\over 2} \xi \to n_{\parallel}-n_{\perp}$, the preceding discussion generalizes to the case in which such a generic medium is slowly 
rotated.  
\vfill\eject

\centerline{\bf 4.~~Ellipticity and polarization axis of the emerging wave}
\bigskip

We proceed next to calculate the polarization 
parameters of the emerging wave, following 
the exposition of Born and Wolf [9].  Let us write 
$$\eqalign{
X=&\cos\theta  + i\Delta^I_X \simeq \cos\theta  
e^{i\Delta^I_X/\cos\theta}\equiv a_1e^{i\delta_1}~~~,\cr
Y=&\sin \theta  i\Delta^I_Y\simeq \sin \theta  
e^{i\Delta^I_Y/\sin \theta}\equiv a_2e^{i\delta_2}~~~,\cr
}\eqno(11a)$$
with  $\Delta^I_{X,Y}$ first order small quantities that can be read   
off from Eq.~(10b).  
Then the auxiliary angle $\alpha$ of Born and Wolf is given by 
$$\tan \alpha ={a_2 \over a_1}= \tan \theta ~~~,\eqno(11b)$$
which implies that $\alpha =\theta$ up to second order corrections, and  
so to first order there is no rotation of the polarization axis of the beam. 
The second auxiliary angle 
$\chi$ of Born and Wolf is given by $\sin 2 \chi=\sin 2\alpha   \sin \delta$;  
since $\delta = \delta_2-\delta_1
=\Delta^I_Y/\sin \theta - \Delta^I_X/\cos \theta$ is first order small, 
to leading order 
(and allowing the ellipticity to carry a sign) we have 
$$\eqalign{
\Psi=&{\rm Ellipticity}=\tan\chi \simeq  \chi \simeq {1\over 2} \delta 
\sin 2\alpha  \simeq {1\over 2} \delta \sin 2\theta ~~~\cr
=&\cos \theta \Delta^I_Y-\sin \theta \Delta^I_X~~~.\cr
}\eqno(11c)$$ 
{}From the values of $\Delta^I_{X,Y}$ obtained from Eq.~(10b), we then find 
$$\Psi=- {3\over 4} \xi \bar \omega L  \sin2(\theta-\Omega t)~~~,\eqno(11d)$$
which using $\bar \omega = 2 \pi/\lambda$ and $\xi=2 \Delta n/3$ is 
equivalent to 
$$\Psi=-{\pi L \over \lambda} \Delta n \sin2(\theta-\Omega t)~~~.\eqno(11e)$$
Thus, to first order in small quantities, the ellipticity is exactly 
what would be calculated by assuming the magnetic field to be frozen at 
its instantaneous position during its traversal by the laser beam, using 
the formula for ellipticity calculated from the vacuum birefringence in a 
static magnetic field.  
\bigskip
\centerline{\bf 5.~~Perturbation method for a 
rotating, time-dependent magnetic field}
\bigskip
The simple form of the  final answer of Eq.~(10b) suggests there should 
be a simpler derivation than the wave matching procedure employed in 
Sec. 3.  We show in this section that the same answer can be obtained by 
a perturbation theory approach, which while not yielding the 
structure of the propagation modes in the magnetic field region, has the 
advantage that it extends to the case when the magnetic field strength 
(that is, the coefficient $\xi$) is also time dependent. Substituting the  
general Ansatz 
$$\eqalign{
\vec e=&E_1(z,t) \hat B + E_2(z,t) \hat z \times \hat B~~~,\cr   
\vec b=&B_1(z,t) \hat B + B_2(z,t) \hat z \times \hat B ~~~,\cr   
}\eqno(12a) $$
into Eq.~(1a), and using Eq.~(1b) but now allowing $\xi$ to have a $t$ and  
$z$ dependence, we get the coupled equations 
$$\eqalign{
\partial_zE_1(z,t)=&-\Omega B_1(z,t) -\partial_tB_2(z,t)~~~,\cr
-\partial_zE_2(z,t)=&-\partial_tB_1(z,t)+\Omega B_2(z,t)~~~,\cr
(1-6\xi)\partial_zB_1(z,t)=&(1+5\xi)\Omega E_1(z,t)
+(1-2\xi)\partial_tE_2(z,t)-2 \partial_t\xi E_2(z,t)
+6\partial_z\xi B_1(z,t)~~~,\cr 
-(1-2\xi)\partial_zB_2(t)=&(1+5\xi)\partial_tE_1(z,t)-(1-2\xi)\Omega E_2(z,t)
+5 \partial_t \xi E_1(z,t)
-2\partial_z\xi B_2(z,t)~~~.\cr
}\eqno(12b)$$
We can solve 
these equations by making a perturbation expansion in powers of $\xi$, 
by writing $E_{1,2}=E_{1,2}^{(0)}+E_{1,2}^{(1)}+...$, and similarly for 
$B_{1,2}$, with the fields with subscript $(0)$ the incident wave  
components on rotating bases, 
$$\eqalign{
E_1^{(0)}=&e^{i\bar \omega (z-t)} \cos(\theta-\Omega t)~~~,\cr 
E_2^{(0)}=&e^{i\bar \omega (z-t)} \sin(\theta-\Omega t)~~~,\cr 
B_1^{(0)}=&-e^{i\bar \omega (z-t)} \sin(\theta-\Omega t)~~~,\cr 
B_2^{(0)}=&e^{i\bar \omega (z-t)} \cos(\theta-\Omega t)~~~,\cr 
}\eqno(13a)$$
and with the fields with superscript $(1)$ perturbations proportional to 
$\xi$.  The zeroth order fields are exact solutions of Eq.~(12b) when 
$\xi$ is set to zero, but $\Omega$ is kept non-zero. Substituting  
the perturbation expansion into Eq.~(12), differentiating 
$\partial_zE^{(1)}_{1,2}$ 
with respect to $z$ to get $\partial^2_z E^{(1)}_{1,2}$, and using 
Eq.~(12b) to eliminate cross derivatives  
$\partial_t\partial_zB^{(1)}_{1,2}$ in terms 
of $E^{(1)}_{1,2}$, we get inhomogeneous wave equations for the first order 
perturbations,  
$$\eqalign{
(\partial_z^2-\partial_t^2)E_{1,2}^{(1)}=&I_{1,2}~~~,\cr            
I_1=&7\xi \partial_t^2E_1^{(0)} +12\partial_t\xi \partial_t E_1^{(0)}
+5 \partial^2_t\xi E_1^{(0)}
-2\partial_z\xi\partial_t E_1^{(0)}-2\partial_z\partial_t\xi E_1^{(0)}~~~,\cr
I_2=&4\xi \partial_t^2 E_2^{(0)}+2\partial_t \xi \partial_t E_2^{(0)}
 -2\partial^2_t \xi E_2^{(0)}
-6\partial_z\xi\partial_t E_2^{(0)}-6\partial_z\partial_t\xi E_2^{(0)}~~~,\cr
}\eqno(13b)$$                         
where we have dropped terms of order $\xi\Omega$ 
(but have kept $\Omega$ in 
the phases of the zeroth order solution, where it multiplies $t$, which 
can be arbitrarily large). 
Using the Green's function $G(z,t)$ for the one-dimensional 
wave equation with outgoing wave boundary conditions, 
$$\eqalign{
G(z,t)=&-{1\over 2} \theta(t-|z|)~~~,\cr
(\partial_z^2-\partial_t^2)G(z,t)=&\delta(z)\delta(t)~~~,\cr
}\eqno(13c)$$ 
we can solve Eq.~(13b) to give 
$$
E_{1,2}^{(1)}(z,t)=\int_0^L dz^{\prime} \int_{-\infty}^{\infty} dt^{\prime} 
G(z-z^{\prime},t-t^{\prime})
I_{1,2}(z^{\prime},t^{\prime})~~~.\eqno(14)$$
A similar solution can be obtained for the perturbations 
$B_{1,2}^{(1)}$, giving the general first order corrections to the fields,    
even when the magnetic field strength is $t$ and $z$ dependent. 

Let us now consider the case when the field strength is only weakly varying 
in time, 
so that $\partial_t\xi /\xi$ is a small parameter.  Then to first order 
in small quantities, the terms in Eq.~(13b) involving time derivatives of 
$\xi$  as well as explicit powers of $\Omega$ can be dropped, 
so that $I_{1,2}$ reduce to 
$$\eqalign{
I_1=&e^{i\bar \omega (z-t)} \cos(\theta-\Omega t)
[-7 \xi\bar \omega^2+2i\bar\omega  \partial_z \xi]  ~~~,\cr
I_2=&  e^{i\bar \omega (z-t)} \sin(\theta-\Omega t)
[-4 \xi\bar \omega^2+6i\bar\omega  \partial_z \xi]~~~.\cr
}\eqno(15a)$$
When the magnetic field is uniform (although possibly weakly time 
dependent) in the interval $0\leq z \leq L$, and zero elsewhere, one has 
$\partial_z\xi=\bar \xi(t) [\delta(z)-\delta(z-L)]$, with $\partial_t \bar 
\xi(t)$ second order small.  Then 
substituting Eq.~(15a) into Eq.~(14), taking $z=L$, so that  
$|z-z^{\prime}|=L-z^{\prime}$ on the support of the integrand, and 
regarding terms of order 
$\Omega/\bar \omega$ and $\Omega L$ as small, we can evaluate the  
integrals over $z^{\prime},t^{\prime}$ to get the first order fields. For 
the contribution from the 
terms $-(7,4)\xi\bar \omega^2$ in the square brackets in Eq.~(15a), which  
have support on the 
interval $0\leq z^{\prime} \leq L$, we use
$$\eqalign{
&\int_0^L dz^{\prime}\int_{-\infty}^{\infty} dt^{\prime} 
\theta(t-t^{\prime}-L+z^{\prime})  
e^{i\bar \omega (z^{\prime}-t^{\prime})\pm i\Omega t^{\prime} } \cr
&=\int_0^L dz^{\prime}\int_{-\infty}^{t-L+z^{\prime} } dt^{\prime} 
e^{i\bar \omega (z^{\prime}-t^{\prime})\pm i\Omega t^{\prime} } \cr
&\simeq\int_0^L dz^{\prime}{i\over \bar \omega} e^{i\bar \omega (L-t)    
\pm i \Omega (t-L+z^{\prime})}\cr
&\simeq {iL\over \bar \omega}   e^{i\bar \omega (L-t)\pm i \Omega t}~~~,\cr     
}\eqno(15b)$$ 
while for the contribution from the terms $(2,6)i\bar \omega \partial_z\xi$  
in the square brackets in Eq.~(15a), which have support on the  
field boundaries  at $z^{\prime}=0,L$, we use
$$\eqalign{
&\int dz^{\prime} \int dt^{\prime}  \theta(t-t^{\prime}-L+z^{\prime})  
\bar \xi(t^{\prime})  [\delta(z^{\prime})-\delta(z^{\prime}-L)]
e^{i\bar \omega (z^{\prime}-t^{\prime})\pm i\Omega t^{\prime} } \cr
&=\int_{-\infty}^t du e^{i \bar \omega (L-u) \pm i\Omega u}
[\bar \xi(u-L) e^{\mp i\Omega L} - \bar \xi(u)]\cr
&={\rm O}(\partial_u\bar \xi,\,\bar \xi \Omega L)\simeq 0~~~.\cr
}\eqno(15c)$$
Adding back the zeroth order fields, we get for the fields at $z=L$,   
$$\eqalign{
E_1|_{z=L}=&e^{i\bar \omega(z-t)} \cos(\theta-\Omega t)(1+{7\over 2}i\bar 
\omega \xi L)~~~,\cr
E_2|_{z=L}=&e^{i\bar \omega(z-t)} \sin(\theta-\Omega t)(1+{4\over 2}i\bar 
\omega \xi L)~~~,\cr
}\eqno(16a)$$
which when transformed back to the fixed bases by  
$$E_1|_{z=L}(\cos \Omega t,\sin \Omega t, 0) +E_2|_{z=L} (-\sin \Omega t,\cos \Omega t,0)
=e^{i\bar \omega(z-t)} (X,Y)~~~,\eqno(16b)$$ 
give 
$$\eqalign{
X=&\cos \theta (1+{11\over 4}i \bar \omega L \xi) 
+{3\over 4} i\bar \omega L \xi \cos(2\Omega t-\theta)~~~,\cr
Y=&\sin \theta (1+{11\over 4}i \bar \omega L \xi) 
+{3\over 4} i\bar \omega L \xi \sin(2\Omega t-\theta)~~~,\cr
}\eqno(16c)$$
in agreement with our earlier result of Eq.~(10b).  
\bigskip
\centerline{\bf 6.~~Discussion}
\bigskip
Returning to Eq.~(5b), we see that the form of the propagation eigenmodes 
in the magnetic field region is governed by the behavior of the 
square root term in $\sigma_{\pm}$, that is, by 
$\big((2\Omega/\omega)^2+(3\xi/2)^2\big)^{1\over 2}$, with distinctly 
different behaviors  
in the small $\Omega$ regime $2\Omega/\omega<<3\xi/2$, and the 
large $\Omega$ regime $2\Omega/\omega>>3\xi/2$.  In the former, the 
refractive indices are $n_+ \simeq 1+(7/2)\xi$ and 
$n_- \simeq 1+2 \xi$, as in the case of a non-rotating magnetic field. 
In the latter, the refractive indices are $n_{\pm} \simeq  
1+(11/4) \xi \pm \Omega/ \omega$. This change in character of the  
eigenmodes does not show up in the transmitted wave polarization 
parameters, however, because all  
dependence on the square root cancels between the various terms in 
Eq.~(11a). The simple results that we get for the polarization parameters   
are what one would immediately get by assuming that the dependence on 
$\xi$ and $\Omega$ should be analytic around the origin in these parameters, 
since then the fact that effects of the magnetic field region vanish at 
$\xi=0$ implies that there can be no first order term proportional to 
$\Omega$; the rotation dependence must first enter at second order through 
terms proportional to $\xi \Omega$.  Similarly, dependences on a small time 
rate of change of the magnitude of the magnetic field, which are relevant 
for experiments in which a non-rotating but ramping magnetic field is used, 
must also appear only in second order terms in small parameters.  

We note in conclusion that 
putting in the numbers characterizing the PVLAS experiment, even though 
their magnetic field rotates only a fraction of a revolution per second, 
their parameters yield $\Omega L=0.7 \times 10^{-8}$, 
$2\Omega/\bar \omega \simeq 2 \times 10^{-15}$, and $3 \xi/2 \simeq 
10^{-22}$. So although all of these are small,  PVLAS is in fact 
operating in the large $\Omega$ regime in terms 
of the behavior of the square root and the propagation eigenmodes in the 
rotating field region. Nonetheless, as we have shown, the formulas they  
have used for the ellipticity and polarization axis orientation 
of the emerging wave are correct.  
\bigskip
\centerline{\bf Acknowledgments}
This work was supported in part by the Department of Energy under
Grant \#DE--FG02--90ER40542.  I wish to thank Ra\'ul Rabad\'an for bringing 
the paper of ref [6] to my attention, and participants in the ``Axions at 
the Institute for Advanced Study'' workshop for stimulating discussions.  
After the initial draft of this paper was posted to the arXiv, I benefited 
from informative email correspondence with Carlo Rizzo and Holger Gies, 
and from helpful referee comments.  
\medskip
\noindent{\it Added note:} After this work was completed, we learned 
of related work by Biswas and Melnikov [10], whose results agree with 
ours.      
\bigskip
\bigskip
\bigskip
\centerline{\bf References}
\bigskip
\noindent
[1]  Heisenberg W and Euler H (1936) {\it Z. Physik} {\bf 98} 714 
\hfill\break
\bigskip 
\noindent
[2]   Schwinger J (1951) {\it Phys. Rev.} {\bf 82} 664 \hfill \break
\bigskip
\noindent
[3]  Toll J (1952) unpublished Princeton dissertation \hfill \break
\bigskip
\noindent
[4]  Adler S L (1971) {\it Ann. Phys.} {\bf 67} 599 \hfill\break  
\bigskip
\noindent
[5]  Zavattini E et al. (2006) {\it Phys. Rev. Lett.} {\bf 96} 110406; 
Bakalov D et al. (1994) {\it Nucl. Phys. B Proc. Suppl.} 
{\bf 35} 180; Bakalov D et al. (1998) {\it Quantum Semiclass. Opt.} 
{\bf 10} 239; Bakalov D et al. (1998) {\it Hyperfine Interact.} 
{\bf 114} 103.  \hfill\break
\bigskip
\noindent
[6] Tito Mendon\c ca J, Dias de Deus J and Castelo Ferreira P (2006)  
{\it Phys. Rev. Lett.} {\bf 97} 100403 \hfill\break
\bigskip
\noindent
[7]  Tito Mendon\c ca J, Dias de Deus J and Castelo Ferreira P (2006)  
arXiv:hep-ph/0609311\hfill\break
\bigskip
\noindent
[8] Problems with the calculation of ref [6] 
were first pointed out by Holger Gies (unpublished memorandum, July, 2006),  
and problems with ref [6] and ref [7] were noted 
independently by the author (unpublished memorandum, October, 2006)   
\hfill\break
\bigskip
\noindent
[9]  Born M and Wolf E  (1964) {\it Principles of Optics} 
(New York:Macmillan); for a concise exposition with application to 
PVLAS, see \hfill\break
Adler, S L (2006) http://www.sns.ias.edu/$\sim$adler/talks.html 
\hfill\break 
\bigskip
\noindent
[10] Biswas S and Melnikov K (2006) arXiv: hep-ph/0611345
\bye
\bigskip
\noindent
\bigskip
\noindent
\bigskip
\noindent
\bigskip
\noindent
\bigskip
\noindent
\bigskip
\noindent
\bigskip
\noindent
\bigskip
\noindent
\bigskip
\noindent
\bigskip
\noindent
\bigskip
\noindent
\bigskip
\noindent
\vfill
\eject
\centerline{\bf Figure Captions}
\noindent
Fig. 1  
\bigskip
\noindent
Fig. 2  
\bigskip
\noindent
Fig. 3  
\bigskip
\noindent
Fig. 4  
\bigskip
\noindent
Fig. 5  
\bigskip
\noindent
Fig. 6  
\bigskip
\noindent
Fig. 7  
\bigskip
\bye